\documentclass[aps,preprint,showpacs,amsmath,amssymb]{revtex4}

\usepackage{graphicx}
\usepackage{amssymb}
\usepackage{bm}
\usepackage{subfigure}

\begin{document}
\setcounter{page}{1}

\title{Dynamics of relaxation, decoherence and entropy of a qubit in anisotropic photonic crystals}

\author{Jing-Nuo \surname{Wu$^{1}$}}
\author{Hung-Kuang \surname{Chen$^{2}$}}
\author{Wen-Feng \surname{Hsieh$^{3}$}}
\email {wfhsieh@mail.nctu.edu.tw}
\thanks{FAX:+886-3-5716631}
\author{Szu-Cheng \surname{Cheng$^{1}$}}
\email {sccheng@faculty.pccu.edu.tw}
\thanks{FAX:+886-2-28610577}

\affiliation{$^{1}$Department of Physics, Chinese Culture University, Taipei, Taiwan, R. O. C.}
\affiliation{$^{2}$Department of Electronic Engineering, National Chin-Yi University of Technology, Taichung, Taiwan, R. O. C.}
\affiliation{$^{3}$Department of Photonics and Institute of Electro-Optical Engineering, National Chiao Tung University, Hsinchu, Taiwan, R. O. C.}
\date[]{Received \today }

\begin{abstract}
We study the quantum dynamics of relaxation, decoherence and
entropy of a qubit embedded in an anisotropic photonic crystal (PhC)
through fractional calculus. These quantum measurements are
investigated by analytically solving the fractional Langevin
equation. The qubit with frequency lying inside the photonic band gap (PBG) exhibits the preserving behavior of energy, coherence and information amount through the steady values of excited-state probability, polarization oscillation and von Neumann entropy. This preservation does not exist in the Markovian system with qubit frequency lying outside the PBG region. These accurate results are based on the appropriate mathematical method of fractional calculus and reasonable inference of physical phenomena.  
\end{abstract}
\pacs{03.67.-a, 42.70.Qs, 42.50.Dv}
\maketitle

\section{INTRODUCTION}
A central tenet in quantum computing is to create and manipulate
qubits in a controllable way. Many novel
methods have been proposed to generate controllable entangled states
through the surrounding environment of the qubits \cite{LMazzola09,
SManiscalco08, Bellomo08}. When a qubit made of a two-level atom is
embedded inside a reservoir with structured photon density of state
(DOS), the evolution dynamics of its states is affected by the
environment and exhibits non-Markovian behavior \cite{Jorgensen11,
SJohn94, Wu2010}. One of structured reservoirs is three-dimensional
(3D) photonic crystals (PhCs). A qubit embedded in a PhC with photon
DOS containing a cut-off below the band edge frequency can form a photon-atom
bound state and show strong non-Markovian dynamics which provides
the basis of manipulating a qubit in this structured reservoir.

PhCs are ordered crystals with periodic variation of refractive index and show
photonic band gaps (PBGs) in their band structures. For a practical 3D PhC,
the dispersion relation near the PBG edge exhibits a directional-dependent property
and thus has been expressed as a vector form by the effective-mass
approximation \cite{NVats02}. This 3D material with the anisotropic
band structure possesses a DOS proportional
to $\sqrt{\omega-\omega_{c}}\Theta(\omega-\omega_{c})$, where $\omega_{c}$ is the band edge frequency and the
Heaviside step function $\Theta(\omega-\omega_{c})$ characterizes the cut-off behavior. Recently, the strong effect of
this structured DOS on the spontaneous emission (SE) dynamics is observed in the
experiment of quantum dots embedded in the diamond-based PhC
lattice \cite{Jorgensen11}. The strong light-matter interaction
resulting from the structured DOS leads to the non-Markovian SE
dynamics including non-exponential decay and formation of the photon-atom
bound state. The decay rate of this non-Markovian SE can be varied by a
factor larger than $10$ through controlling the lattice parameters of this 3D structure.
This modification of SE rate will lead
to the change of quantum coherence of the system. The related coherence effects include
the electromagnetically induced transparency (EIT) \cite{Harris97}, slow propagation of
light \cite{Hau01}, and nonlinear effects \cite{Harris89}. These coherent effects protect
the embedded qubits from suffering the destruction such as entanglement sudden death (ESD) by the environment
and make the PhCs become the promising candidate for the reservoir of the quantum
information systems \cite{Dutton, Saffman10}

However, the previous theoretical studies about the SE dynamics of the PhCs
through the conventional Laplace transform method delivered the physical
results with some fallacious predictions which are inconsistent with
experiments \cite{SJohn94, YYang03A}. These qualitatively accurate results have
been used frequently as a model for studying the dynamics of quantum
measurements for qubits embedded in PhCs \cite{Bellomo08, Amri09}.
The misleading quantitative conclusions made by these studies would lead to the
inaccurate strategies to manipulate qubits in PhCs.

As a qubit is embedded in a PhC reservoir, the dynamics of this qubit is related to the delay Green function which is a measure of the PhC reservoir memory on the qubit. This delay Green function, also named memory kernel, depends strongly on the DOS of the PhC reservoir. The memory kernel of the PhC reservoir with a cut-off DOS exhibits the long-time memory effect, which leads to the non-Markovian memory dynamics of SE. This non-Markovian dynamics can be solved accurately through fractional calculus which have been applied in studying the SE dynamics of PhCs \cite{SCCheng09, Wu2010}. The study for the isotropic PhC system through fractional calculus shows that the SE dynamics exhibits no photon-atom bound state as the qubit frequency lies outside the PBG region. This result is consistent with the experimental result that no prolonged lifetime effect exists for the emission peak lying outside the PBG region \cite{SCCheng09, MFujita05}. The study about SE dynamics from anisotropic PhCs by the use of fractional calculus shows that the anisotropic property of the PhC reservoir enhances the decaying behavior of the qubit which agrees with the experimental observation that emission spectrum appears only in the PhC system with strong anisotropic property for the emission frequency lying inside the PBG region \cite{Wu2010, MBarth06}. These accurate results of the SE dynamics by the use of fractional calculus reveal that this mathematical method is appropriate for the PhC systems with non-Markovian dynamics. We therefore use fractional calculus to investigate the dynamics of quantum measurements for a qubit embedded in an anisotropic PhC.

The controllable characteristic of SE dynamics from PhCs provides the basis of many novel methods of manipulating qubits \cite{Bellomo08, Entezar09, Amri09, Jorgensen11}. For example, in Ref. \onlinecite{Bellomo08} Bellomo \textit{et} \textit{al}. suggested high values of entanglement trapping between two independent qubits can be achieved through choosing the isotropic PhC materials as reservoir. This achievement of entanglement preservation is based on the direct link between the SE dynamics of the single qubit and the entanglement dynamics. Whenever the qubit forms a bound state with the emitted photon through the PhC reservoir, entanglement trapping follows. However, because they followed the predictions of previous studies with some quantitative fallaciousness \cite{SJohn90, SJohn94}, they predicted the unphysical entanglement trapping occurred for the qubit frequency lying outside the PBG region. This problem also happened in Ref. \onlinecite{Entezar09} where Entezar explored the entanglement dynamics of a single qubit and its radiation field near the PBG edge. He predicted that there exists a steady-state entanglement between the qubit and radiation field even when the qubit frequency lies outside the PBG region. This prediction was made by supposing that the system would form the photon bound state as the qubit frequency lies outside the PBG region. This supposed fact conflicts the experimental results that photon bound states exist only for the region deep inside the PBG but not for the outside PBG region\cite{MBarth06}. Studying accurately about how a single qubit evolves in the PhC reservoir provides the reliable basis of strategies to manipulate qubits in PhCs. We thus aim at finding out the accurate dynamics of a single qubit in a PhC reservoir to provide the reliable basis of manipulating qubits in the PhC environment.

In this paper, dynamics of quantum measurements including the relaxation, decoherence and von Neumann entropy of a qubit embedded in a PhC with one-band anisotropic model are investigated by the use of fractional calculus. The system exhibits distinct dynamics as the qubit frequency lies inside and outside PBG region. Dynamics of the non-Markovian system with qubit frequency lying outside PBG region exhibits quickly suppression of relaxation and decoherence which implies the preservation of qubit energy and coherence. The coherent correlation between the qubit and the PhC reservoir leads to the preservation of information amount stored in the qubit through the steady entropy with non-zero value. 

The paper is organized as follows. In Sec. II, the theoretical model for the system of a qubit in an anisotropic PhC is depicted through the Hamiltonian and the wave function. In use of the fractional calculus, we express the kinetic equation of the qubit as a fractional Langevin equation and solve this equation analytically to construct the reduced density matrix of the qubit. Dynamics of relaxation, polarization, decoherence and von Neumann entropy are discussed based on this matrix. Quantum operations of amplitude damping, polarization damping and multistage computations on the single qubit in the non-Markovian and Markovian systems are analyzed according to these discussions. Finally, we summarize our results in Sec. III.

\section{Dynamics of a qubit in an anisotropic PhC}
When we consider a system consisting of a qubit embedded inside a PhC reservoir shown in Fig. 1, the Hamiltonian can be expressed as
\begin{equation}
H=\frac{1}{2}\hbar\omega_{10}(\left|1\right\rangle\left\langle 1\right|-\left|0\right\rangle\left\langle 0\right|)+\sum_{\vec{k}}{\hbar\omega_{\vec{k}}a_{\vec{k}}^{+}a_{\vec{k}}}+i\hbar\sum_{\vec{k}}{g_{\vec{k}}(\left|0\right\rangle\left\langle 1\right|\otimes a_{\vec{k}}^{+}+{h.c.})}.
\end{equation}
Here $\omega_{10}$ is the qubit transition frequency between the excited state $|{1}\rangle$ and ground state $|{0}\rangle$. The photon mode with frequency $\omega_{\vec{k}}$ is created and annihilated by the operators $a_{\vec{k}}^{+}$ and $a_{\vec{k}}$. Coupling strength associated with the qubit and the photon mode with frequency $\omega_{\vec{k}}$ is specified by $g_{\vec{k}}=\frac{\omega_{10}d_{10}}{\hbar}[{\frac{\hbar}{2\epsilon_{0}\omega_{\vec{k}}V}]}^{\frac{1}{2}}\hat{e}_{\vec{k}}\cdot\hat{u}_d$ with the fixed qubit dipole moment $\vec{d}_{10}=d_{10}\hat{u}_d$, sample volume $V$, dielectric constant $\epsilon_{0}$ and polarization unit vector $\hat{e}_{\vec{k}}$ of the photon mode with frequency $\omega_{\vec{k}}$.

If we use the coordinate ($\theta,\phi$) on the Bloch sphere to parameterize the state of the qubit with $(0,\phi)$ for the excited state and $(\pi,\phi)$ for the ground state, then the initial state of the total system can be written as
\begin{equation}
|\psi(0)\rangle=\left[e^{i\phi_{0}}cos(\frac{\theta_{0}}{2})|1\rangle+sin(\frac{\theta_{0}}{2})|0\rangle\right]\otimes \left|\textbf{0}_{\vec{k}}\right\rangle+\sum_{\vec{k}}{C_{\vec{k}}(0)|0\rangle\otimes \left|\textbf{1}_{\vec{k}}\right\rangle}
\end{equation}
with the initial coordinate ($\theta_{0},\phi_{0}$) on the Bloch sphere. Here the vacuum state and one-excitation state of the PhC reservoir are expressed as $\left|\textbf{0}_{\vec{k}}\right\rangle=\left|0_{1},0_{2},...,0_{\vec{k}},0_{\vec{k}+1},...\right\rangle$ and $\left|\textbf{1}_{\vec{k}}\right\rangle=\left|0_{1},0_{2},...0_{\vec{k}-1},1_{\vec{k}},0_{\vec{k}+1},...\right\rangle$, respectively. We assume $C_{\vec{k}}(0)=0$ for no initial correlation between the qubit and reservoir. As the system evolves, the quantum state of the system in the single photon sector can be written as
\begin{equation}
|\psi(t)\rangle=\left[u_{p}(t)e^{i\phi_{0}}cos(\frac{\theta_{0}}{2})|1\rangle+u_{d}(t)sin(\frac{\theta_{0}}{2})|0\rangle\right]\otimes \left|\textbf{0}_{\vec{k}}\right\rangle+\sum_{\vec{k}}{C_{\vec{k}}(t)|0\rangle\otimes \left|\textbf{1}_{\vec{k}}\right\rangle}
\end{equation}
with initial condition $u_{p}(0)=1$, $u_{d}(0)=1$ and $C_{\vec{k}}(0)=0$. Here $u_{p}(t)$ and $u_{d}(t)$ stand for the excited-state and ground-state probability amplitudes of the qubit with the vacuum state of the PhC reservoir while $C_{\vec{k}}(t)$ for the qubit in its ground state with one photon in the reservoir. 
This quantum state can also be expressed as $|\psi(t)\rangle=\left\{e^{i\phi (t)}cos[\frac{\theta (t)}{2}]|1\rangle+sin[\frac{\theta (t)}{2}]|0\rangle\right\}\otimes  \left|\textbf{0}_{\vec{k}}\right\rangle+\sum_{\vec{k}}{C_{\vec{k}}(t)|0\rangle\otimes \left|\textbf{1}_{\vec{k}}\right\rangle}$ with the time-dependent Bloch parameters [$\theta (t),\phi (t)$] when we discuss the change of the qubit's state on the Bloch sphere.

By substituting the quantum state in Eq. (3) into the time-dependent Schr$\ddot{o}$dinger equation, we obtain the equations of motion for the amplitudes as
\begin{equation}
\frac{d}{dt}u_{p}(t)=-\frac{1}{cos(\theta_{0}/2)}\sum_{\vec{k}}{g_{\vec{k}}C_{\vec{k}}(t)e^{-i\Omega_{\vec{k}}t}},
\end{equation}
\begin{equation}
\frac{d}{dt}C_{\vec{k}}(t)=g_{\vec{k}}u_{p}(t)cos(\theta_{0}/2)e^{i\Omega_{\vec{k}}t},
\end{equation}
and
\begin{equation}
\frac{d}{dt}u_{d}(t)=0
\end{equation}
with detuning frequency $\Omega_{\vec{k}}=\omega_{\vec{k}}-\omega_{10}$. The last equation of motion yields $u_{d}(t)=u_{d}(0)=1$ meaning that the ground-state probability with vacuum photon mode will not evolve with time, i.e., a time constant. The other two equations can be combined as
\begin{equation}
\frac{d}{dt}u_{p}(t)=-\int_{0}^{t}{G(t-\tau)u_{p}(\tau)d\tau}
\end{equation}
with the memory kernel
$G(t-\tau)=\sum_{\vec{k}}{g_{\vec{k}}^{2}e^{-i\Omega_{\vec{k}}(t-\tau)}}$. This equation of motion reveals that the future state of the qubit is related to the memory of the reservoir in its previous state through the memory kernel. As the qubit is put in free space, the memory kernel has the form of a Dirac delta function $G(t-\tau)\propto\delta(t-\tau)$ corresponding to continuous photon DOS. In this case, the reservoir manifests its memory effect only at an instant time $\tau=t$ which leads to the excited amplitude of the qubit decaying exponentially with time. This Markovian result in free space manifests that the qubit loses all memory of its past and decays quickly to its ground state.

For the anisotropic PhC reservoir discussed here, the memory kernel manifests its memory effect within the entire interval (0,t) through $G(t-\tau)=\frac{\beta^{1/2}/f^{3/2}}{\sqrt{\pi}(t-\tau)^{3/2}}e^{-i[3\pi/4-\delta(t-\tau)]}$  with the coupling constant $\beta^{1/2}=(\omega^{2}_{10}d^{2}_{10}\sqrt{\omega_{c}})/(16\pi\epsilon_{0}\hbar c^{3})$ and the detuning frequency $\delta=\omega_{10}-\omega_{c}$ of the qubit transition frequency $\omega_{10}$ from the band edge frequency $\omega_{c}$ \cite{Wu2010}. This memory kernel is derived from the anisotropic dispersion relation of a practical three-dimensional PhC. Near the band edge frequency $\omega_{c}$, the dispersion relation has a vector form and could be expressed by the effective-mass approximation as \cite{SJohn87} $\omega_{\vec{k}}\approx\omega_{c}+A\left(\vec{k}-\vec{k}_{c}\right)^{2}$ where the curvature $A\cong f\omega_{c}/k^{2}_{c}=fc^{2}/\omega_{c}$ signifies its different values in different directions through the scaling factor $f$. This anisotropic dispersion relation leads to the memory kernel expressed by the cut-off photon DOS $\rho(\omega)$ through $G(t-\tau)=\frac{\omega^{2}_{10}d^{2}_{10}}{4\epsilon_{0}\hbar}\int^{\infty}_{0}d\omega\frac{\rho(\omega)}{\omega}e^{-i(\omega-\omega_{10})(t-\tau)}$ with $\rho(\omega)=\frac{1}{4\pi^{2}}\sqrt{\frac{\omega-\omega_{c}}{A^{3}}}\Theta(\omega-\omega_{c})$ and the Heaviside step function $\Theta(\omega-\omega_{c})$ charactering the cut-off behavior. With this memory kernel for the anisotropic PhC reservoir, the equation of motion in Eq. (7) becomes
\begin{equation}
\frac{d}{dt}u_{p}(t)=-\frac{\beta^{1/2}e^{i3\pi/4}}{\sqrt{\pi}f^{3/2}}\int^{t}_{0}\frac{u_{p}(\tau)e^{i\delta(t-\tau)}}{(t-\tau)^{3/2}}d\tau.
\end{equation}
This equation can be further simplified through making the transformation $u_{p}(t)=e^{i\delta t}U_{p}(t)$ which gives
\begin{equation}
\frac{d}{dt}U_{p}(t)+i\delta U_{p}(t)=\frac{\beta^{1/2}e^{i\pi/4}}{\sqrt{\pi}f^{3/2}}\int^{t}_{0}\frac{U_{p}(\tau)}{(t-\tau)^{3/2}}d\tau.
\end{equation}
Conventionally, this integro-differential equation is solved through Laplace transform which leads to the fractal phenomenon of the system from this memory kernel in Laplace image \cite{RRNigmatullin92}. This fractal phenomenon would result in the stochastic nature of the dynamical behavior of the system which appears as the non-Markovian dynamics. This non-Markovian dynamics can be solved accurately through fractional calculus which we have shown its appropriateness by comparing the obtained results with the experimental ones and finding their consistence \cite{Wu2010, SCCheng09}.

In the following, we use fractional calculus to solve the non-Markovian dynamics of the system analytically. Especially, the fractional time derivative, one of the operators of fractional calculus, is used to express the integral term of the kinetic equation (9). In the well-known Riemann-Liouvile definition, the fractional time derivative operator $d^{\nu}/dt^{\nu}$ is expressed as \cite{IPodlubny99}
\begin{equation}
\frac{d^{\nu}}{dt^{\nu}}f(t)=\frac{1}{\Gamma (n-\nu)}\frac{d^{n}}{dt^{n}}\int^{t}_{a}\frac{f(\tau)}{(t-\tau)^{\nu-n+1}}d\tau
\end{equation}
for $n-1\leq\nu<n$ and $\Gamma (x)$ being the Gamma function. Through expressing the right-hand-side term of Eq. (9) as the fractional time derivative operator with order $n=1/2$ and applying the appropriate fractional operations to this kinetic equation, we arrive at the fractional kinetic equation of this qubit-reservoir interacting system as 
\begin{equation}
\frac{d^{1/2}}{dt^{1/2}}U_{p}(t)+i\delta\frac{d^{-1/2}}{dt^{-1/2}} U_{p}(t)+\frac{2\beta^{1/2}e^{i\pi/4}}{f^{3/2}}U_{p}(t)=\frac{t^{-1/2}}{\sqrt{\pi}}.
\end{equation}
The fractional time derivative in this fractional differential equation indicates a subordinated stochastic process directing to a stable probability distribution \cite{AAStanislavsky04}. The evolution equation governing the future of the qubit is expressed as the form of a \textsl{fractional Langevin equation} because of its interacting with the PhC reservoir.
Our aim of studying the quantum dynamics of relaxation, decoherence and entropy for the qubit system can be achieved by solving this equation through the Laplace transform for the fractional operators. The basic formula used here is
\begin{equation}
\textsl{L}\left\{\frac{d^{\nu}}{dt^{\nu}}f(t)\right\}\equiv\int^{\infty}_{0}e^{-st}\frac{d^{\nu}}{dt^{\nu}}f(t)dt=s^{\nu}\textsl{L}\left\{f(t)\right\}-\sum^{n-1}_{m=0}s^{m}\left[\frac{d^{\nu-m-1}}{dt^{\nu-m-1}}f(t)\right]_{t=0}
\end{equation}
with $s$ denoting the Laplace variable. The procedure of performing the Laplace transform on the fractional Langevin equation leads to
\begin{equation}
\tilde{U}_{p}(s)=\frac{1}{s+i\delta+2\beta^{1/2}e^{i\pi/4}s^{1/2}/f^{3/2}}
\end{equation}
where $\tilde{U}_{p}(s)$ is the Laplace transform of ${U}_{p}(t)$. This equation can be further expressed as a sum  of partial fractions as the roots of the indicial equation $Y^{2}+2\beta^{1/2}e^{i\pi/4}Y/f^{3/2}+i\delta=0$ are found with the variable $s^{1/2}$ having been converted into $Y$. As $\beta/f^{3}\neq\delta$, we have the different roots of the indicial equation which leads to the partial-fractional form of $\tilde{U_{p}}(s)=\left[\frac{1}{(\sqrt{s}-Y_{1})}-\frac{1}{(\sqrt{s}-Y_{2})}\right]\frac{1}{(Y_{1}-Y_{2})}$ with $Y_{1}=e^{i\pi/4}\left(-\frac{\beta^{1/2}}{f^{3/2}}+\sqrt{\frac{\beta}{f^{3}}-\delta}\right)$ and $Y_{2}=e^{i\pi/4}\left(-\frac{\beta^{1/2}}{f^{3/2}}-\sqrt{\frac{\beta}{f^{3}}-\delta}\right)$. For $\beta/f^{3}=\delta$, the indicial equation has degenerate root leading to $\tilde{U_{p}}(s)=1/\left(\sqrt{s}+\frac{\beta^{1/2}e^{i\pi/4}}{f^{3/2}}\right)^{2}$. 

The inverse Laplace transform of these partial-fractional forms can not be found in the conventional mathematical table because the power of the variable $s$ is not an integer. For the fractional power of $s$, we can find its inverse Laplace transform in the book of fractional calculus \cite{KSMiller93} as  $\textsl{L}^{-1}\left\{\frac{1}{(s^{\nu}-a)}\right\}=\sum^{q}_{j=1}a^{j-1}E_{t}(j\nu-1,a^{q})$ and  $\textsl{L}^{-1}\left\{\frac{1}{(s^{\nu}-a)^{2}}\right\}=\sum^{q}_{j=1}\sum^{q}_{m=1}a^{j+m-2}\left\{tE_{t}\left((j+m)\nu-2,a^{q}\right)-\left[(j+m)\nu-2\right]E_{t}\left((j+m)\nu-1,a^{q}\right)\right\}$ with $\nu=1,1/2,1/3,...$ and $q=1,2,...,1/\nu$. Here we name the two-parameter function $E_{t}(\alpha,a)$ as the fractional exponential function with variable $t$, order $\alpha$ and constant $a$. It is defined as the fractional derivative of an ordinary exponential function $E_{t}(\alpha,a)\equiv\frac{d^{-\alpha}}{dt^{-\alpha}}e^{at}=t^{\alpha}\sum^{\infty}_{n=0}\frac{(at)^{n}}{\Gamma (\alpha+n+1)}$ with the derivative formula $\frac{d^{\mu}}{dt^{\mu}}E_{t}(\alpha,a)=E_{t}(\alpha-\mu,a)$. The functional form of this fractional exponential function $E_{t}(\alpha,a)$ will not be changed after being performed derivative operator with fractional or integral order. A linear combination of the fractional exponential functions is a potential solution of the fractional differential equation. By applying these inverse Laplace transforms to the partial-fractional forms of $\tilde{U}_{p}(s)$, we obtain the analytical solution of the fractional differential equation [11] as
\begin{equation}
U_{p}(t)=\frac{1}{2e^{i\pi/4}\sqrt{\beta/f^{3}-\delta}}\times\left[Y^{2}_{1}E_{t}(1/2,Y^{2}_{1})-Y^{2}_{2}E_{t}(1/2,Y^{2}_{2})+Y_{1}e^{Y^{2}_{1}t}-Y_{2}e^{Y^{2}_{2}t}\right],
\end{equation}
for $\beta/f^{3}\neq\delta$; and
\begin{equation}
U_{p}(t)=-2\frac{\beta^{3/2}e^{i3\pi/4}}{f^{9/2}}tE_{t}(\frac{1}{2},i\beta/f^{3})-\frac{\beta^{1/2}e^{i\pi/4}}{f^{3/2}}E_{t}(\frac{1}{2},i\beta/f^{3})
+(1+\frac{2it\beta}{f^{3}})e^{i\beta t/f^{3}}-2\frac{\beta^{1/2}e^{i\pi/4}}{f^{3/2}\sqrt{\pi}}t^{1/2}
\end{equation}
for $\beta/f^{3}=\delta$. Here we have applied the recursion relation of the fractional exponential function $E_{t}(\nu,a)=aE_{t}(\nu+1,a)+\frac{t^{\nu}}{\Gamma(\nu+1)}$ to these analytical expressions.
The time evolution of the wave function is thus obtained as
\begin{equation}
|\psi(t)\rangle=\left[u_{p}(t)e^{i\phi_{0}}cos(\frac{\theta_{0}}{2})|1\rangle+u_{d}(t)sin(\frac{\theta_{0}}{2})|0\rangle\right]\otimes \left|\textbf{0}_{\vec{k}}\right\rangle+\sum_{\vec{k}}{C_{\vec{k}}(t)|0\rangle\otimes \left|\textbf{1}_{\vec{k}}\right\rangle}
\end{equation}
with $u_{p}(t)=e^{i\delta t}U_{p}(t)$ expressed through Eq. (14) and (15), $u_{d}(t)=u_{d}(0)=1$ and $\sum_{\vec{k}}\left|C_{\vec{k}}(t)\right|^{2}=1-\left[u_{p}(t)cos(\frac{\theta_{0}}{2})\right]^{2}-\left[sin(\frac{\theta_{0}}{2})\right]^2$.

By definition, the reduced density matrix of the qubit can be directly obtained from this wave function through tracing over the reservoir degrees of freedom. It gives   
\begin{equation}
\hat{\rho}(t)=\bordermatrix{ \cr  &  \cr
  & \left|u_{p}(t)\right|^{2}cos^{2}(\frac{\theta_{0}}{2}) &  \frac{1}{2}u^{*}_{p}(t)e^{-i\phi_{0}}sin(\theta_{0}) \cr   &  \frac{1}{2}u_{p}(t)e^{i\phi_{0}}sin(\theta_{0})  & 1-\left|u_{p}(t)\right|^{2}cos^{2}(\frac{\theta_{0}}{2}) \cr}\equiv\bordermatrix{ \cr  &  \cr
  & \rho_{11}(t) &  \rho_{10}(t) \cr   & \rho_{01}(t)  & \rho_{00}(t) \cr}
\end{equation}
with the initial one $\hat{\rho}(0)=\bordermatrix{ \cr  &  \cr
  & cos^{2}(\frac{\theta_{0}}{2}) &  \frac{1}{2}e^{i\phi_{0}}sin(\theta_{0}) \cr   & \frac{1}{2}e^{-i\phi_{0}}sin(\theta_{0})  & sin^{2}(\frac{\theta_{0}}{2}) \cr}$. The elements in this matrix are associated with the polarization and probabilities of the qubit. In the following, we will study the quantum measurements of relaxation, decoherence and von Neumann entropy based on these elements.

\subsection{Relaxation rate}
As the excited qubit spontaneously emits a photon, the qubit undergoes a quantum operation of amplitude damping, the excited-state amplitude of the qubit being reduced. This amplitude damping operation associated with the energy dissipation of the qubit leads to the flow of energy from the qubit into the environment through the emitted photon. How the energy flows from the initially excited qubit can be observed from the excited-state probability of the qubit $P(t)$ through $P(t)=\left|u_{p}(t)\right|^{2}cos^{2}(\frac{\theta_{0}}{2})=\left|U_{p}(t)\right|^{2}$ ($\theta_{0}=0$) shown in Fig. 2(a). For the qubit frequency lying inside PBG [$\delta/\beta=(\omega_{10}-\omega_{c})/\beta<0$], the probability dynamics exhibits decay and oscillatory behavior before reaching a non-zero steady-state value. The energy flow from the qubit to the PhC reservoir is paused through this non-zero steady-state value. The excited-state energy of the qubit is preserved by keeping the emitted photon bound with the qubit. Bound states with larger steady-state values have less energy flow from the excited qubit to the reservoir. The rate of this energy flow can be measured through the relaxation rate $\Gamma_{relax.}(t)=-\frac{\dot{\rho}_{11}(t)}{\rho_{11}(t)}=-\left[\frac{\dot{U}_{p}(t)}{U_{p}(t)}+\frac{\dot{U}^{*}_{p}(t)}{U^{*}_{p}(t)}\right]$ with $U^{*}_{p}(t)$ being the complex conjugate of $U_{p}(t)$ in Eqs. (14) and (15). The dynamics of the relaxation rate in Fig. 2(b) from this expression decays quickly and becomes zero thereafter for the qubit frequency lying inside PBG ($\delta/\beta<0$). Combining with the dynamical behavior of excited-state probability, we find that the qubit loses partial of its energy in the very beginning period of time and then preserves the remaining energy in the end. The quickly stop of losing energy results from the formation of the bound state between the qubit and the reflected photon. The remaining energy is preserved through this bound state in the end. Energy localization around the qubit by the bound state leads to the non-decaying dynamics of excited-state probability in Fig. 2(a) and quickly decaying of relaxation rate in Fig. 2(b). Especially, as the qubit frequency is detuned deep inside the PBG region ($\delta/\beta=-5$ and $-10$), the relaxation rate exhibits negative values in the initial period of time. This negative-relaxation-rate phenomenon reveals that the qubit can regain its energy in the PhC reservoir.

On the other hand, for the qubit frequency lying outside PBG region [$\delta/\beta=(\omega_{10}-\omega_{c})/\beta>0$], the relaxation rate decays slowly and remains a constant of non-zero value at longer time. It corresponds to that the excited qubit loses almost its energy in the initial period of time with $\beta t<2$ shown in Fig. 2(a). In this Markovian case, the emitted photon flees from the qubit without being re-absorbed so that it carries away all of the qubit's energy to the reservoir.

When we use a Bloch vector in the Bloch sphere to express the qubit state, the amplitude damping operation will perform a transformation on the components of the Bloch vector with the values related to the probability of losing the emitted photon \cite{Nielsen00}. For the Markovian system with qubit frequency lying outside the PBG region, the large-value probability of losing the emitted photon will convert the Bloch vector toward the ground state point on the Bloch sphere. The small-value probability of losing the emitted photon in the non-Markovian system with qubit frequency lying inside PBG leads to the Bloch vector towards the non-zero large surface area near the ground-state point on the Bloch sphere. The contracting effect of the amplitude damping on the Bloch vector leading to the contracted surface area of the Bloch sphere available for the Bloch vector is suppressed in the non-Markovian system.  

\subsection{Decoherence}
As a qubit interacts with the environment, there will randomize the polarization of each qubit. The quantum phase information of a qubit carried by the qubit polarization will thus escape from the qubit into the environment through this randomization of polarization and lead to the quantum decoherence. As the qubit undergoes the polarization damping operation through this quantum decoherence, the off-diagonal elements of the reduced density matrix will decay with time. The completely losing of the quantum coherence of one qubit will lead to the occurrence of ESD in the multiple-qubit system. That is, the time period in which the entanglement could be usefully exploited in the multi-qubit system is limited. This limitation is related to the dynamics of qubit polarization and decoherence. Here we show these dynamics in Fig. 3 through the expressions of qubit polarization $Pz(t)=\frac{1}{2}[\rho_{10}(t)+\rho_{01}(t)]=Re[e^{i\delta t}U_{p}(t)]$ and decoherence rate $\Gamma_{dec.}(t)=-\frac{\dot{\rho}_{10}(t)+\dot{\rho}_{01}(t)}{\rho_{10}(t)+\rho_{01}(t)}=-\left[\frac{\dot{U}_{p}(t)+\dot{U}^{*}_{p}(t)}{U_{p}(t)+U^{*}_{p}(t)}\right]$. 

In Fig. 3(a), the polarization dynamics of the Markovian system with qubit frequency lying outside the PBG region ($\delta/\beta=2$) exhibits fast damping behavior which implies that the qubit loses all its polarization in the initial period of time with $\beta t<4$. This polarization damping results from the elastic scattering by the incoherence photons of the environment. The elastic scattering randomizes the orientation of the qubit's polarization and leads to the decay of the off-diagonal elements of the density matrix. The fast damping of the qubit polarization reveals that the probability of scattering the qubit by the incoherent photons in the reservoir is large in the Markovian system. The continuous photon DOS outside the PBG region [see Fig. 1(c)] provides the free-space-like environment to scatter the qubit. 

In contrast, the polarization dynamics of the non-Markovian system with qubit frequency lying inside PBG ($\delta/\beta<0$) exhibits non-decaying oscillation. The qubit loses partial of its polarization in the very beginning period of time and then preserves the remaining polarization through the steady oscillation. This steady polarization ensures that the probability of scattering the qubit by the incoherent photons is greatly lowered in the non-Markovian system. The inhibited photon DOS of the PhC reservoir inside the PBG region [see Fig. 1(c)] is expected to lower this probability of scattering the qubit and lead to the preservation of phase information of the qubit through the steady polarization. States with larger amplitudes of steady oscillation have less lose of phase information from the qubit to the reservoir. The rate of this lose of phase information can be measured by the decoherence rate $\Gamma_{dec.}(t)$ shown in Fig. 3(b). The decoherence rate of the non-Markovian system decays quickly to zero. Combining the results of Fig. 3(a), we infer that the phase information in the qubit is partially lost in the very beginning period of time and then is preserved through the steady polarization. In this non-Markovian case, the interaction with the PhC reservoir does not result in the total lose of the coherence but in the preservation of the partial coherence instead. This coherence preservation reveals that PhC materials can provide a promising environment for the purpose of information processing because they meet the requirement of engineering the information-processing systems through the greatly minimized quantum decoherence.

Comparing the dynamics of the relaxation rate [Fig. 2(b)] with that of the decoherence rate [Fig. 3(b)], we can find the similarity between these two quantum measurements. This is because both the relaxation and decoherence arise from the effects of spontaneous emission. As the emitted photon flees from the qubit, it carries away both the energy and phase information of the qubit. We compare these two quantum measurements through calculating the ratio of the decoherence rate to the relaxation rate. This ratio exhibits constant at $0.5$ for both the Markovian and non-Markovian regimes. This result shows that the decoherence will proceed much slower than the energy relaxation in this PhC reservoir. That is, the decoherence time $T_{1}$ is twice the time of the energy relaxation $T_{2}$, i.e., $T_{1}=2T_{2}$. This familiar relation between the decoherence time and relaxation time for the generic case of a two-level system was also observed in a transverse reservoir with a Lorentzian spectrum in the infinite-temperature limit \cite{Kofman00,Nielsen00}. When the energy relaxation is completed, almost half of the initial coherence of the qubit will still survive.

\subsection{Von Neumann entropy}
Entropy, a key concept of quantum information theory, measures how much information exists in the state of a physical system. The amount of information will be changed as the state is correlated to the environment. The correlation between the environment and the state will transform the initially pure state into a finally mixed state where the amount of information of the state is changed. We thus can estimate the correlation between the qubit and reservoir through the entropy of the quantum system, i.e., von Neumann entropy. Von Neumann entropy for a quantum state $\hat{\rho}(t)$ is defined as $S(t)=-Tr\left[\hat{\rho}(t) log\hat{\rho}(t)\right]=-\sum_{i}\lambda_{i}log\lambda_{i}$ with $\lambda_{i}$ being the eigenvalues of the matrix $\hat{\rho}(t)$. In this expression, the entropy quantifies the optimal compression for the physical resources required to solve some information problem because in the measurement the eigenvalues $\lambda_{i}$ stand for the expected values and $log\lambda_{i}$ for the probability of obtaining $\lambda_{i}$. That is, von Neumann entropy predicts the upper bound of the information we will gain after we measure a quantum state. This gain of information after measurement corresponds to the information amount of the state before measurement. The value of this entropy stands for the degree of correlation between the qubit and the PhC reservoir. It reaches the maximal value $S_{max}=log d$, $d$ being the dimension of the matrix $\hat{\rho}(t)$, for the qubit's maximally mixed state and zero value for the pure state. With the eigenvalues $\lambda_{\pm}=\frac{1}{2}\left\{1\pm\sqrt{1-4cos^{4}\left(\frac{\theta_{0}}{2}\right)\left[\left|u_{p}(t)\right|^{2}-\left|u_{p}(t)\right|^{4}\right]}\right\}$ of the matrix in Eq. (17), we show the von Neumann entropy in Fig. 4 for the initially excited qubit ($\theta_{0}=0$). The entropy has its minimal value zero at $t=0$ and reaches its maximal value $log 2=0.693$ at the very beginning of time. After a period of time on the order of the decay timescale, the entropy
becomes steady with nonzero value in the non-Markovian system ($\delta/\beta<0$) and zero in the Markovian system ($\delta/\beta=2$). These results show that the initially pure system becomes maximally mixed in the very beginning period of time. As the qubit equilibrates with the PhC reservoir, the system becomes less mixed. In the Markovian system, the qubit recovers its initially pure state as it quickly equilibrates with the reservoir. However, in the non-Markovian system, the coherent correlation with the PhC reservoir leads to the preservation of the mixed state. The greatly lowered probability of the qubit scattering by the incoherent photons in the PBG reservoir preserves the coherence and information amount of the qubit.  

When we perform a multi-stage computations on the available information through the Markovian chain, the data processing inequality of von Neumann entropy, a basic inequality of information theory, states that the information about the output of the computation will decrease with time if the qubit storing the information is correlated to the Markovian environment \cite{Nielsen00}. This statement agrees with the result we obtained here. In the Markovian system, the newly produced mixed state by the correlation between the qubit and reservoir recovers its initially pure state and loses the information stored in the qubit. The informations yielded by the multi-stage computations will thus be independent of each other. On the other hand, in the non-Markovian system, the newly produced mixed state will preserve the information about its output of the previous stage. The memory effect of the PhC reservoir on the qubit's previous state leads to the preservation of the stored information in the mixed-state qubit.  Further computation operations on this mixed state can be used to increase the amount of mutual information between the outputs of the operations and the previous-stage information about the qubit state.

\section{Conclusion}
We have studied the dynamics of quantum measurements of relaxation,
decoherence and von Neumann entropy for a qubit in an
anisotropic PhC. Through applying fractional calculus to solve the
fractional Langevin equation, we obtain the analytical expressions
of these quantum measurements. The quantum operations corresponding
to these quantum measurements act differently in the non-Markovian
and Markovian systems which are differentiated according to the qubit frequency lying inside or outside the PBG region. For the non-Markovian system with qubit frequency lying inside PBG ($\delta/\beta<0$), the relaxation dynamics shows that the qubit can preserve the remaining energy through forming the bound state with the photon. The small-value probability of losing the emitted photon leads to the suppressed contracting effect of the amplitude damping operator on the qubit's vector in the Bloch sphere. Without forming the bound state with the emitted photon, the excited qubit of the Markovian system with frequency lying outside the PBG region ($\delta/\beta>0$) loses all its energy to the reservoir. This large-value probability of losing the emitted photon leads to the Bloch vector of the qubit contracting to the ground-state point on the Bloch sphere. The distinct behavior between the non-Markovian and Markovian systems also exists in the decoherence and entropy dynamics. Steady polarization oscillation and entropy with non-zero value in the non-Markovian system reveals that the greatly lowered probability of scattering the qubit by the incoherent photons in the PhC reservoir leads to the preservation of the coherence and information amount of the qubit. This preservation does not exist in the Markovian system with continuous photon DOS outside the PBG region where the PhC reservoir provides the free-space-like environment with large probability of scattering the qubit. Scattering by the incoherent photons leads to fast damping of qubit polarization and losing the information stored in the qubit. These different results from those of previous studies are based on the appropriate mathematical method of fractional calculus and reasonable inference of physical phenomena.


\begin{acknowledgments}
We would like to gratefully acknowledge partially financial support from the National Science Council (NSC), Taiwan under Contract Nos. NSC 100-2811-M-034-003, NSC 100-2221-E-167-031, NSC 99-2112-M-006-017-MY3, NSC 99-2221-E-009-095-MY3, and NSC-99-2112-M-0034-002-MY3.
\end{acknowledgments}


\begin{figure}
\includegraphics[bb=68 308 530 519, width=12.0cm, totalheight=5.0cm, clip]{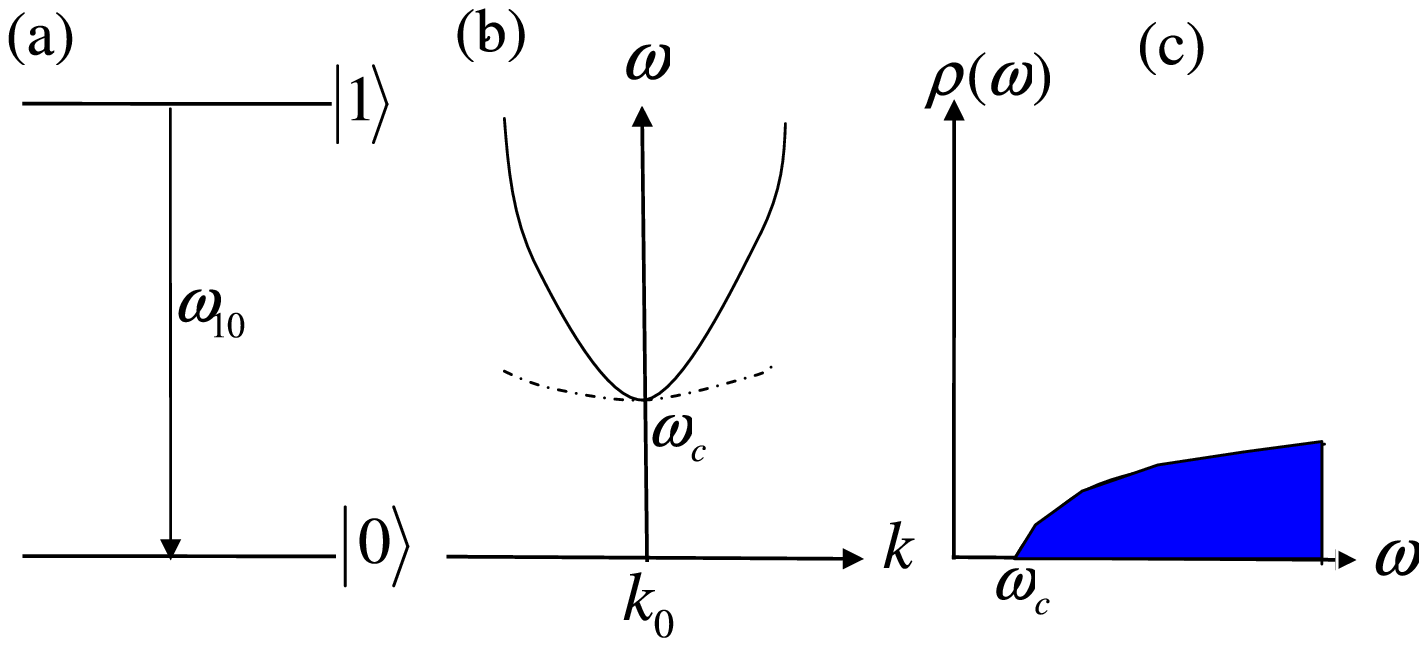}
\caption{(Color online) (a) A qubit with excited state $\left|1\right\rangle$ and ground state $\left|0\right\rangle$. The transition frequency $\omega_{10}$ is nearly resonant with the frequency range of the PhC reservoir. (b) Directional dependent dispersion relation near band edge expressed by the effective-mass approximation with the edge frequency $\omega_{c}$. (c) Photon DOS $\rho(\omega)$ of the anisotropic PhC reservoir exhibiting cut-off photon mode below the edge frequency $\omega_{c}$. }
\end{figure}


\begin{figure}
\subfigure[]{	\includegraphics[bb=5 4 391 293, width=10.0cm,  clip]{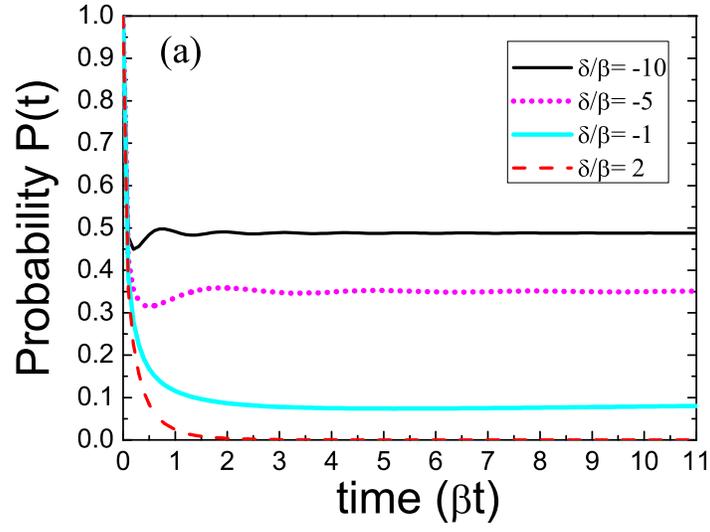}}
\subfigure[]{	\includegraphics[bb=5 4 391 293, width=10.0cm, clip]{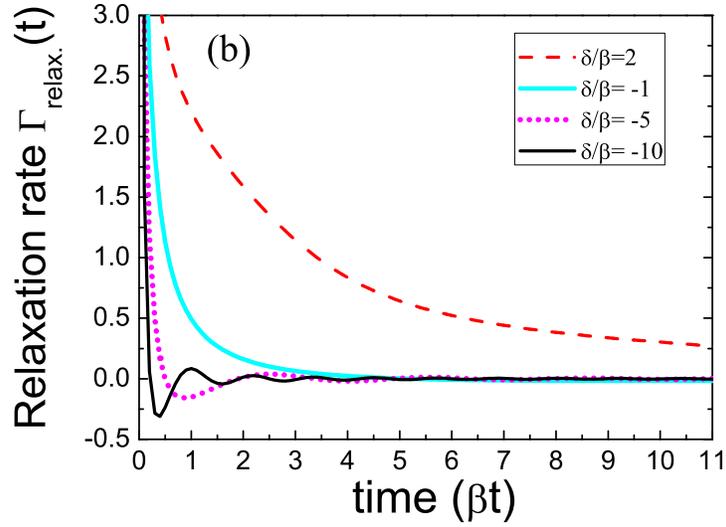}}
\caption{ (Color online) Dynamics of (a) the qubit's excited-state probability $P(t)$ and (b) relaxation rate $\Gamma_{relax.}(t)$ of the qubit with different detuning frequencies $\delta/\beta=(\omega_{10}-\omega_{c})/\beta$ from the band edge frequency $\omega_{c}$ of the PhC reservoir.  }
\end{figure}



\begin{figure}
\subfigure[]{	\includegraphics[bb=5 4 391 293, width=10.0cm, clip]{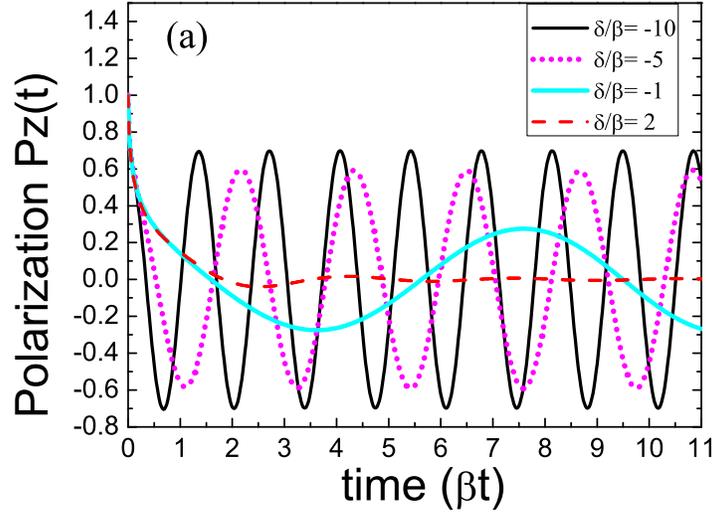}}
\subfigure[]{	\includegraphics[bb=5 4 391 293, width=10.0cm, clip]{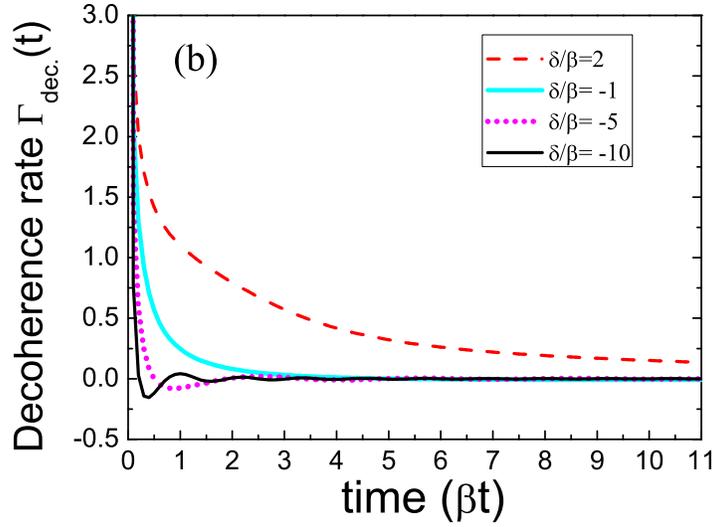}}
\caption{ (Color online) (a) Polarization $Pz(t)=\frac{1}{2}[\rho_{10}(t)+\rho_{01}(t)]$  and (b) Decoherence rate $\Gamma_{dec.}(t)=-\frac{\dot{\rho}_{10}(t)+\dot{\rho}_{01}(t)}{\rho_{10}(t)+\rho_{01}(t)}$ of the qubit with frequency lying inside ($\delta/\beta<0$) and outside ($\delta/\beta=2$) the PBG region . }
\end{figure}


\begin{figure}
\includegraphics[bb=5 4 391 293, width=12.0cm, totalheight=9.5cm, clip]{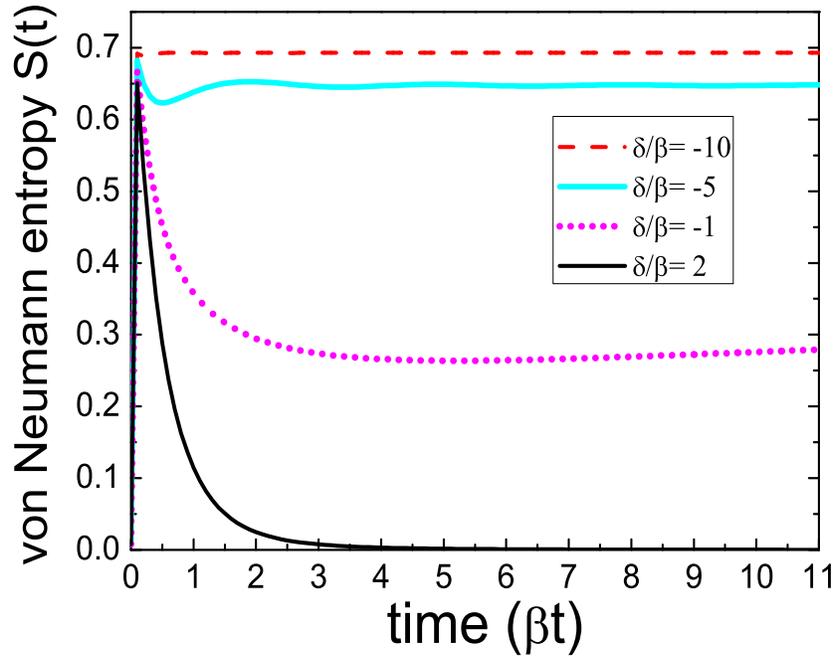}
\caption{(Color online) Von Neumann entropy $S(t)=-Tr\left[\hat{\rho}(t) log\hat{\rho}(t)\right]=-\lambda_{+}log\lambda_{+}-\lambda_{-}log\lambda_{-}$ of the non-Markovian ($\delta/\beta<0$) and Markovian ($\delta/\beta=2$) systems. }
\end{figure}

\end{document}